\documentclass[preprintnumbers,amsmath,amsfonts,superscriptaddress,nofootinbib,aps,prd,twocolumn]{revtex4}
\usepackage[dvips]{color}
\usepackage[dvipdfm]{hyperref}
\usepackage{multirow,slashbox}
\usepackage{graphics}
\usepackage{subfigure}

\newcommand{\newsum}[2]{\overset{#2}{\underset{#1}{\sum}}}

\begin{document} 

\title{New interactions in the dark sector mediated by dark energy} 

\author{Anthony W. Brookfield} 
\email[Email address: ]{php04awb@sheffield.ac.uk} 
\affiliation{Astroparticle Theory and Cosmology Group, Department of
  Applied Mathematics, The University of Sheffield, Hounsfield Road,
  Sheffield S3 7RH, United Kingdom} 
\affiliation{Astroparticle Theory and Cosmology Group, Department of
  Physics and Astronomy, The University of Sheffield, Hounsfield Road,
  Sheffield S3 7RH, United Kingdom} 

\author{Carsten van de Bruck}
\email[Email address: ]{C.vandeBruck@sheffield.ac.uk}
\affiliation{Astroparticle Theory and Cosmology Group, Department of
  Applied Mathematics, The University of Sheffield, Hounsfield Road,
  Sheffield S3 7RH, United Kingdom} 

\author{Lisa M.H. Hall} 
\email[Email address: ]{Lisa.Hall@sheffield.ac.uk}
\affiliation{Astroparticle Theory and Cosmology Group, Department of
Applied Mathematics, The University of Sheffield, Hounsfield Road,
  Sheffield S3 7RH, United Kingdom}
 
 
\begin{abstract} 
Cosmological observations have revealed the existence of a dark matter sector, which 
is commonly assumed to be made up of one particle species only. However, this
sector might be more complicated than we currently believe: there might be more than 
one dark matter species (for example two components of cold dark matter or a mixture of hot and cold dark matter) and there may be 
new interactions between these particles. In this paper we study the possibility of multiple dark matter 
species and interactions mediated by a dark energy field. We study both 
the background and the perturbation evolution in these scenarios. We find that 
the background evolution of a system of multiple dark matter particles (with constant couplings)
mimics a single fluid with a time-varying coupling parameter. However, this is no longer 
true on the perturbative level. We study the case of attractive and repulsive forces as 
well as a mixture of cold and hot dark matter particles. 
\end{abstract}

\maketitle

\section{Introduction}

Cosmological observations of the largest structures in the universe, 
the anisotropies in the cosmic microwave background (CMB) radiation and 
Type 1a supernovae suggest that normal matter contributes only $4\%$
of the total energy budget of the universe \cite{Spergel:2006hy,Seljak:2006bg,Tegmark:2006az}. 
The majority of energy and matter is made up of cold dark matter (CDM, $21\%$), 
which clusters on scales of galaxies and clusters of galaxies, and a much 
more homogeneous distributed dark energy component ($75\%$). 
Whereas dark matter candidates are well motivated within particle theories beyond 
the standard model, it is fair to say that the discovery of the accelerated expansion of the universe 
came as a surprise and that it will be much harder to interpret within particle physics theories. 
Candidates for dark energy include the cosmological constant (e.g. \cite{Carroll:2000fy}), scalar fields 
(see e.g. \cite{Wetterich:1987fm,Peebles:1987ek,Wetterich:1994bg,Zlatev:1998tr,Binetruy:1998rz,Amendola:1999er,
Copeland:2000vh,Copeland:1997et,Brax:1999gp,Brax:1999yv,Ferreira:1997hj,Tocchini-Valentini:2001ty,
Fardon:2003eh,Khoury:2003rn,Farrar:2003uw,Brax:2004qh,Brookfield:2005td}) or vector fields (see e.g. 
\cite{ArmendarizPicon:2004pm,Wei:2006tn,Boehmer:2007qa}). When analysing the 
data it is usually assumed that General Relativity is 
the correct theory of gravity at the largest visible scales. Of course, dark energy might signal 
a breakdown of General Relativity itself, a possibility which has been recently studied in detail (see 
e.g. \cite{Carroll:2003wy,Carroll:2004de,Navarro:2005da,Deffayet:2001pu,Freese:2002sq,Easson:2004fq,Nojiri:2003ft,
Chiba:2003ir,Brookfield:2006mq,Song:2006ej,Bean:2006up,Nojiri:2007as,Amendola:2006kh,Faulkner:2006ub,Appleby:2007vb,
Sawicki:2007tf,Amendola:2006we} and references therein for recent work). In this paper, however, 
we will assume that dark energy can be attributed to a slow-rolling scalar field, interacting with 
other matter forms present in the universe, most notably with CDM and hot dark matter 
(HDM, in the form of massive neutrinos for example). 

Since the nature of dark energy remains unknown, it is important to
study new effects resulting from dark energy and its possible
interactions with the rest of the world. Whereas new forces between
normal matter particles are heavily constrained by observations
(e.g. in the solar system and gravitational experiments on Earth),
this is not the case for neutrinos and dark matter particles. At the
moment, constraints on new forces between neutrinos or dark matter can
only come from cosmological observations. Additionally, it is usually
assumed that there is only one type of dark matter; this is well
motivated by supersymmetrical models which predict that dark matter is
the lightest supersymmetric particle.  However, there might be more
than one dark matter species. Such a possibility has been investigated
previously in \cite{Farrar:2003uw} and can be motivated from string
theory \cite{Gubser:2004uh, Nusser:2004qu}. In this paper, we study the cosmological
implications of a dark sector with 
multiple matter species (CDM, HDM) and consider the possibility
that new forces are mediated between the particles by a
dark energy scalar field. The aim is to study both the background
evolution as well as the evolution of perturbations in the CDM and/or
HDM fluids. We take into account the full evolution of the HDM
component, both for the background as well as on the perturbative
level \cite{Brookfield:2005bz,Ichiki:2007ng}. We will see that there
are new effects associated with new forces in the CDM and/or HDM
sector, and there is much hope that cosmological data will strongly
constrain such interactions. We do not address the question of whether 
the effects of the new interactions are inherent in alternative theories 
of gravity, a possibility which has been pointed out before \cite{Kunz:2006ca}. 
In both coupled quintessence (including models considered here) and modified gravity, the growth rate of density fluctuations will be modified due to the presence of new degrees of freedom. 
This may lead to a degeneracy between these models when only linear perturbations are taken into account.

The new force mediated by the dark energy scalar field is always attractive for particles of the same 
kind. However, if the couplings for the individual species (to be defined in the next section) 
have opposite signs, the force between different forms of matter will be repulsive. For the cases studied in 
this paper the force mediated by dark energy is always long-ranged. As a result, the total force acting 
on the dark sector particles (i.e. the sum of the gravitational force and the new force) 
can be smaller than the gravitational force itself or even repulsive, depending on the coupling. In addition, 
we will see that when the couplings have opposite signs, the expansion of the universe
in the matter dominated era can behave as if there is no new interaction between the particles. Cosmological 
structure formation, however, is affected by the presence of the new forces. 

When studying concrete examples, we will assume that there are only two dark species 
and one dark energy scalar field. The paper is organized as follows. In the next Section, 
we describe the background evolution for three typical models. In Section 3, we study the 
perturbations in these models, calculating both CMB anisotropy and matter power spectra. 
We conclude in Section 4.

\section{Background Evolution}
\label{sec:background}
We consider a system of $N$ species of matter coupled to a single
scalar field $\phi$, which we assume will be responsible for the
late-time accelerated expansion of the universe.  We assume a flat,
homogeneous Friedmann-Robertson-Walker universe with line--element
given by
\begin{equation} 
ds^2=a(\tau)^2\left(-d\tau^2+\delta_{ij}dx^i dx^j \right).
\end{equation}
The Einstein equations, describing the evolution of the scale factor
$a(\tau)$, are given by
\begin{eqnarray}
H^2&\equiv& \left(\frac{\dot{a}}{a}\right)^2 = \frac{8\pi G a^2}{3}\rho,\\
\dot H &=& -\frac{4\pi G a^2}{3}\left(\rho + 3p\right),
\end{eqnarray}
where $\rho(\tau)$ and $p(\tau)$ are the total energy density and
pressure respectively, and overdots represent differentiation with
respect to conformal time $\tau$.  In the following we set $8 \pi G
=1$.

As is customary, we define $\beta_i$ as the strength of the coupling
between the $i^{\rm{th}}$ matter species and the scalar field, which
is given by
\begin{equation}
\beta_i\equiv \frac{d \ln m_i(\phi)}{d\phi},
\end{equation}
where $m_i$ is the particle mass of species $i$.
Conservation of energy momentum for each of the coupled fluids requires that 
\begin{equation}
T^\mu_{\gamma;\mu}=\beta\phi_{,\gamma}T_{~\alpha}^{\alpha}
\label{eq:tmumu}
\end{equation}
which yields
\begin{equation}
\dot{\rho_i}+3 H (\rho_i+p_i)=\beta_i\dot{\phi}(\rho_i-3p_i)
\label{eq:rhodot}
\end{equation}
in the Robertson-Walker spacetime for each of the individual coupled
matter species.  The evolution of the scalar field is described by the
Klein--Gordon equation, which in the presence of matter couplings is
given by
\begin{equation}
\ddot{\phi}+2 H \dot{\phi} + a^2 \frac{dV}{d\phi}=-a^2 \sum_{i=1}^N\beta_i\left(\rho_i-3p_i\right).
\label{eq:kgeqn}
\end{equation}

In
order to investigate the late--time attractor solutions of this
system, we perform the usual transformation of variables
\begin{align*}
x&=\frac{\dot{\phi}}{H \sqrt{6}},  &y&=\frac{1}{H}\sqrt{\frac{V}{3}} \\
z&=\frac{1}{H}\sqrt{\frac{\rho_\gamma}{3}},  &u_i&=\frac{1}{H}\sqrt{\frac{\rho_i}{3}} \\
v&=\frac{1}{H}\sqrt{\frac{\rho_{\mathrm b}}{3}},
\end{align*}
with 
\begin{eqnarray}
\label{friedconst}
x^2+y^2+z^2+\newsum{i=1}{N} u_i^2+v^2= 1.
\end{eqnarray}
In these equations, $\rho_\gamma$ and $\rho_b$ are the energy density of radiation and baryons respectively.
Note that in the following we
assume that $\beta_i$ is constant and $u_i$ is non--relativistic (which is valid for late times).
We also use the Friedmann constraint (Eqn.~(\ref{friedconst})) to eliminate $v$ and consider a general quintessence potential with 
\begin{equation}
\lambda\equiv  -\frac{1}{V}\frac{dV}{d\phi}.
\end{equation}

Differentiation with respect to $\alpha=\ln a$ yields the following
set of equations,
\begin{eqnarray}
H'&=&-\frac{3}{2}H\left[1+x^2-y^2+\frac13z^2\right] \label{eq:Heqn}\\
x'&=&\sqrt{\frac{3}{2}}\lambda y^2-\sum_{i=1}^N \sqrt{\frac{3}{2}}\beta_i u_i^2-x\left(3+\frac{H'}{H}\right)\\
y'&=& -\sqrt{\frac{3}{2}}\lambda x y
-y\frac{H'}{H}\\
z'&=&-z\left(2+\frac{H'}{H}\right)\\
u'_i&=&\sqrt{\frac{3}{2}} \beta_iu_ix-u_i\left(\frac32+\frac{H'}{H}\right).
\label{eq:ueqn}
\end{eqnarray}

It has been shown in  \cite{Tocchini-Valentini:2001ty} that for a single
coupled fluid with an exponential coupling and exponential potential there are
two late time attractor solutions consistent with an accelerating
universe.  The critical point for the first solution (Attractor A) is characterised
by $\Omega_\phi=1$ at late times, and leads to acceleration provided
$\lambda<\sqrt{2}$.  The second critical point (Attractor B) is a scaling solution,
with $\Omega_\phi$ dependent upon the choice of $\beta$ and
$\lambda$.  This solution results in acceleration if $\lambda<2\beta$.
For full details we refer to \cite{Tocchini-Valentini:2001ty}.

We find that introducing multiple coupled fluids does not affect this
result because at late times the system reduces to the single fluid
case as the highly coupled fluids dilute at a faster rate.  Indeed,
the only non--trivial additional critical point which arises due to
the addition of multiple coupled species is characterised by
\begin{equation}
\label{eq:cp3}
\sum_{i=1}^N \beta_i u_i^2=0,
\end{equation}
with $x=y=z=0$ and $\Omega_\phi=0$.  This solution does not lead to an accelerating
cosmology, but can play an important role as a transient solution
during the matter dominated epoch, as we will discuss later.

In order to obtain an understanding of the background evolution, it
turns out to be useful to combine the coupled matter species into a
single effective fluid, with total energy and pressure given by
\begin{equation}
\rho_c=\sum_{i=1}^N\rho_i ,\,\,\,\,p_c=\sum_{i=1}^Np_i,
\end{equation}
and an effective coupling $\beta_{\rm{eff}}$ given by
\begin{equation}
\beta_{\rm{eff}}\equiv
\frac{\newsum{i=1}{N}\beta_i\left(\rho_i-3p_i\right)}{\rho_c-3p_c}.
\label{eq:betaeff}
\end{equation}
We can therefore recast our multi-fluid system of conservation
equations into a single conservation equation for $\rho_c$ with an
effective coupling $\beta_{\rm{eff}}$. We will see later that this is
only possible for the background. It should be further noted that the
effective coupling is a function of both the scalar field and the
energy densities of the coupled fluids, and therefore will in general
vary in time even if the couplings of the individual fluids are
constant. In other words, an observer, studying the background
evolution and assuming a single fluid only, will deduce that the
coupling evolves in time, although there are several species with
constant individual couplings.

\begin{table}[t!]
\begin{tabular}{|c|c|c|c|c|c|c|}  \hline
\multirow{2}{*}{Case} &\multirow{2}{*}{$\lambda$}&\multicolumn{2}{|c|}{Species}&\multicolumn{2}{|c|}{Coupling}&Relative
 Densities\\ \cline{3-7}

    & 	&1&2&	$\beta_1$ & $\beta_2$ &$\left(\frac{\Omega_1}{\Omega_2}\right)_0$\\ \hline\hline
I	  & 	$\sqrt{\frac{3}{2}}$ 	&CDM&CDM&	$4.0$ & $0.1$ &$5.2\times10^{-4}$	\\ \hline
II	  & 	$\sqrt{\frac{3}{2}}$ 	&CDM&CDM&	$-1.5$ & $1$ & $0.28$	\\ \hline
III	  & 	$2$ 	&HDM&CDM&	$5.8$ & $-0.1$ & $0.56$	\\ \hline

\end{tabular}
\caption{Cases considered in this paper: Two cases (I, II) with two
  species of CDM and one case (III) with CDM and HDM.  For all cases
  $\phi_{\rm initial}=0.1$ and we consider a flat universe with cosmological parameters: $h=0.7$, $\Omega_b^{(0)} h^2 = 0.022$ and $\Omega_{\rm CDM}^{(0)} h^2=0.12$ for the coupled CDM species.  
}
\label{tab:setup}
\end{table}

We now consider the effect on the cosmological background of coupling
multiple fluids to the scalar field. For simplicity, we assume that
there are two species of coupled matter, and that the form of
the coupling between the coupled fluids and the scalar field is of 
exponential form for concreteness:
\begin{eqnarray}
m_i(\phi)=m_i^{(*)}e^{\beta_i\phi}.
\end{eqnarray}
We also take a typical quintessence 
exponential potential for the scalar field, namely 
\begin{equation}
V(\phi)=Me^{-\lambda \phi},
\end{equation}
where $M$ and $\lambda$ are chosen to give late time acceleration
today. 

It is instructive to consider three cases, each distinguished by
either the sign of the coupling or the type of dark matter
considered. The specifications can be found in Table \ref{tab:setup}.
In the first two cases (I,II) we consider two types of CDM, but while
the individual couplings $\beta_1$ and $\beta_2$ are both
positive in Case I, the couplings have opposite signs in
Case II. For both of these cases we choose $\lambda$ and $\beta_i$ such
that the system approaches the accelerated late-time attractor with
$\Omega_\phi=1$ (Attractor A). 
A scenario with mass-varying neutrinos (as a single coupled species) in which the system approaches Attractor A has been studied in \cite{Brookfield:2005td,Brookfield:2005bz}.  As will be apparent later, the chosen parameters lead to a consistent background cosmology but the predicted matter power spectra differ significantly from the $\Lambda$CDM case and are observationally ruled out.  The purpose of this paper, however, is to illuminate the physical effects of these models and the extreme parameter choices reflect this.

It is possible to consider a scenario in which a system with two CDM species reaches the late--time accelerated scaling solution (Attractor B).  
In order to achieve an observationally consistent cosmology, this would require one species to remain sub-dominant until today, when it would begin to scale with dark energy.  
In this case, the mass of such a species would vary greatly, most likely leading it to become relativistic in the past.  
This is the third scenario we consider (Case III), in which a dominant CDM
component and a subdominant HDM component each couple to dark
energy.  

A similar
system, albeit with the CDM uncoupled to dark energy, has recently
been considered in \cite{Amendola:2007yx} in the context of neutrinos
with a growing mass.  
We choose $\lambda$ smaller than in this latter paper.
Additionally, we choose to take couplings with opposite
signs, such that the mass of the HDM particle grows in time.
Note that for the HDM species, we fully consider both the non-relativistic and relativistic
behaviour of the particles, integrating the Fermi-Dirac distribution function over phase space \cite{Ma:1995ey}.

In Cases I \& II the particles are highly non--relativistic. Then,
from eqn. (\ref{eq:rhodot}) and using $p_i = 0$, the evolution of the
energy density of the $i^{th}$ coupled fluid is given by
\begin{equation}
\rho_i=\frac{\rho_{i}^{(*)}e^{\beta_i\phi}}{a^3}=\frac{\rho_{i}^{(0)}e^{\beta_i(\phi-\phi_0)}}{a^3},
\end{equation}
where $\rho_{i}^{(0)}$ is the energy density of species $i$ at the present epoch.
With this choice of couplings and two CDM species, an observer assuming a single fluid would detect a time-dependent coupling, which would take the form:
\begin{eqnarray}
\beta_{\rm eff}(\phi) = \frac{Ae^{B\phi}+C}{1+A}
\label{eq:singlebetaeff}
\end{eqnarray} 
where $A=\beta_1\rho_1^{(*)}/\rho_2^{(*)}$, $B=\beta_1-\beta_2$, $C=\beta_2$ and are all constant.

\begin{figure}[t] 
\begin{center}
\scalebox{0.5}{\includegraphics{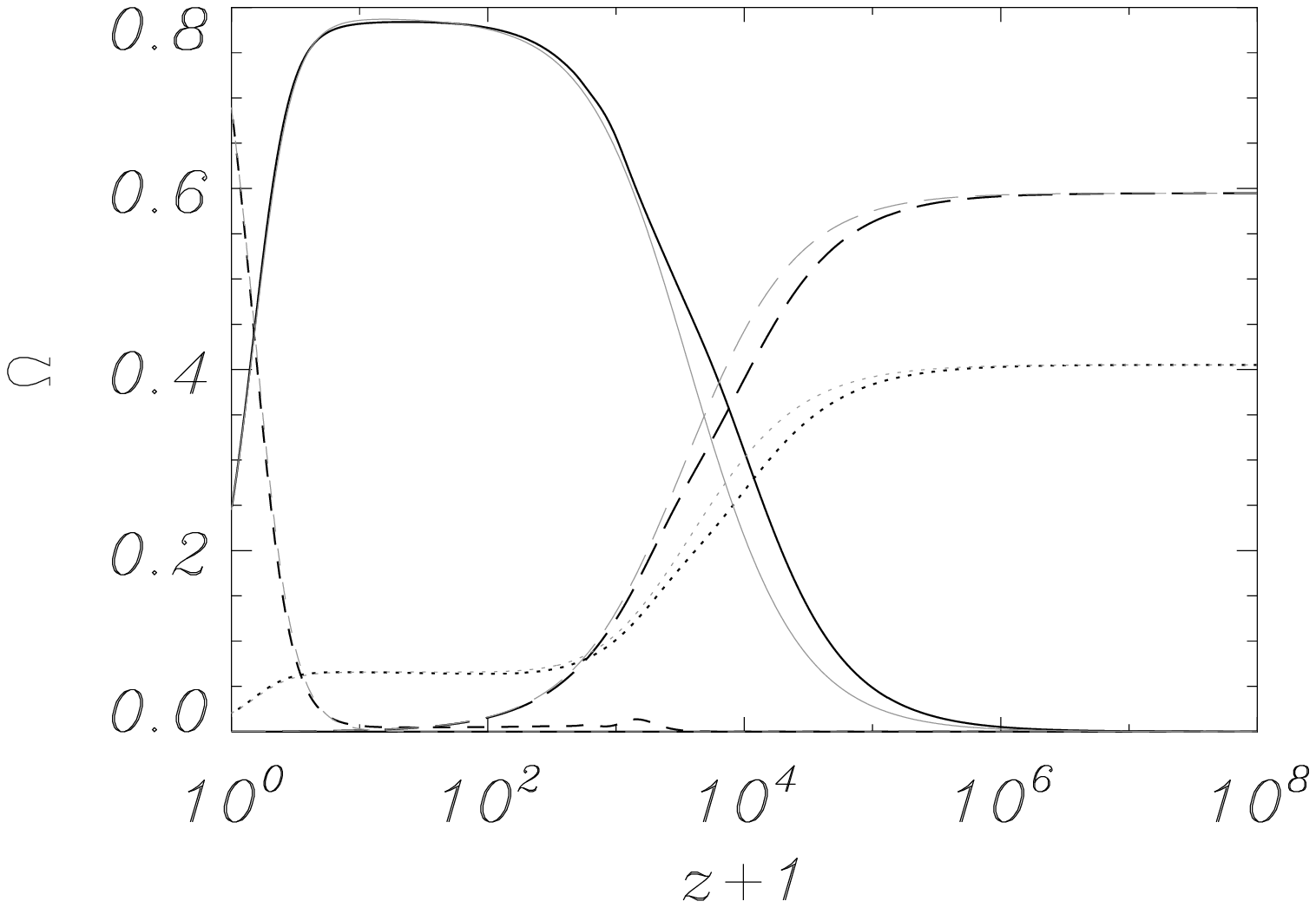}}
\scalebox{0.5}{\includegraphics{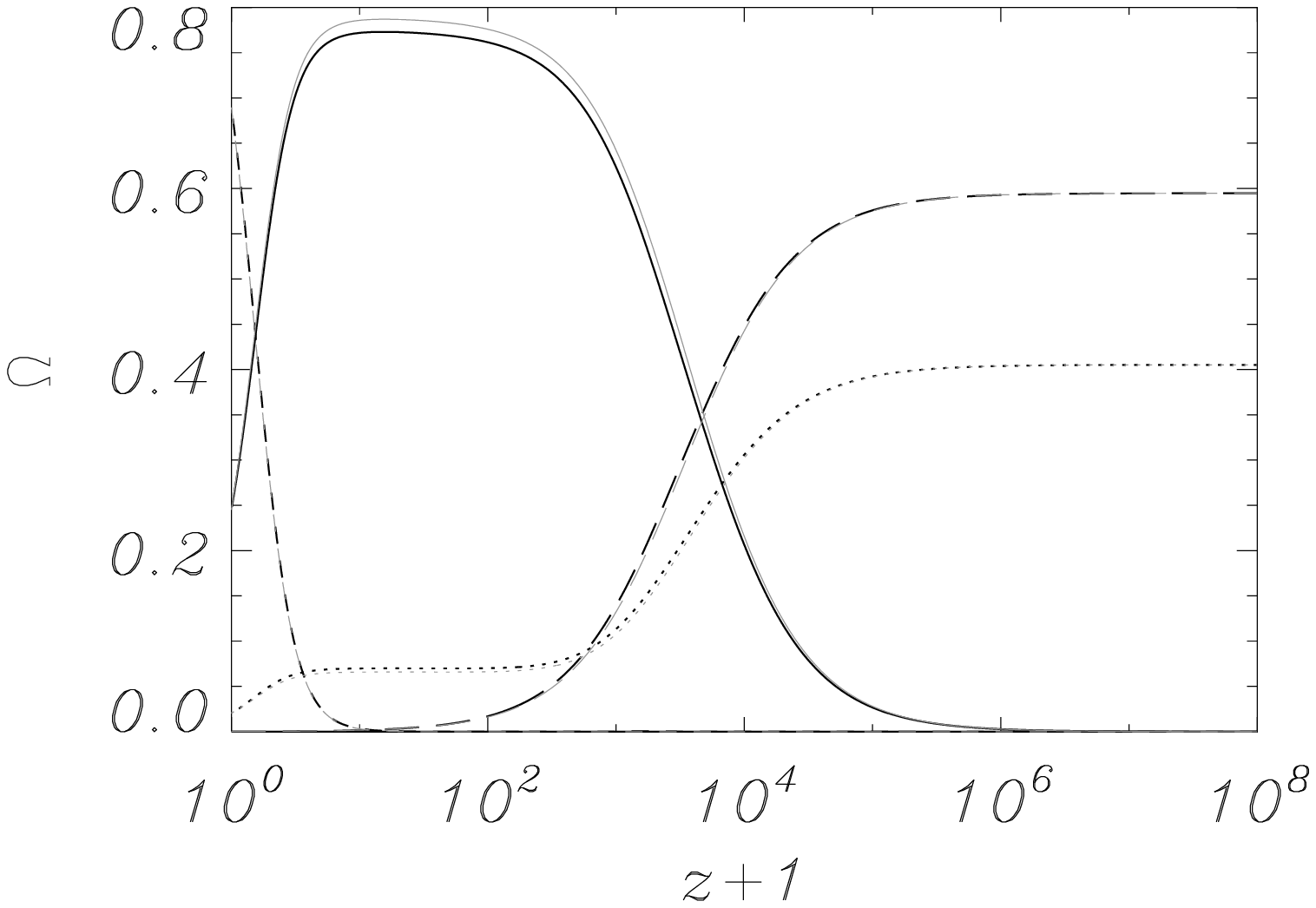}}
\scalebox{0.5}{\includegraphics{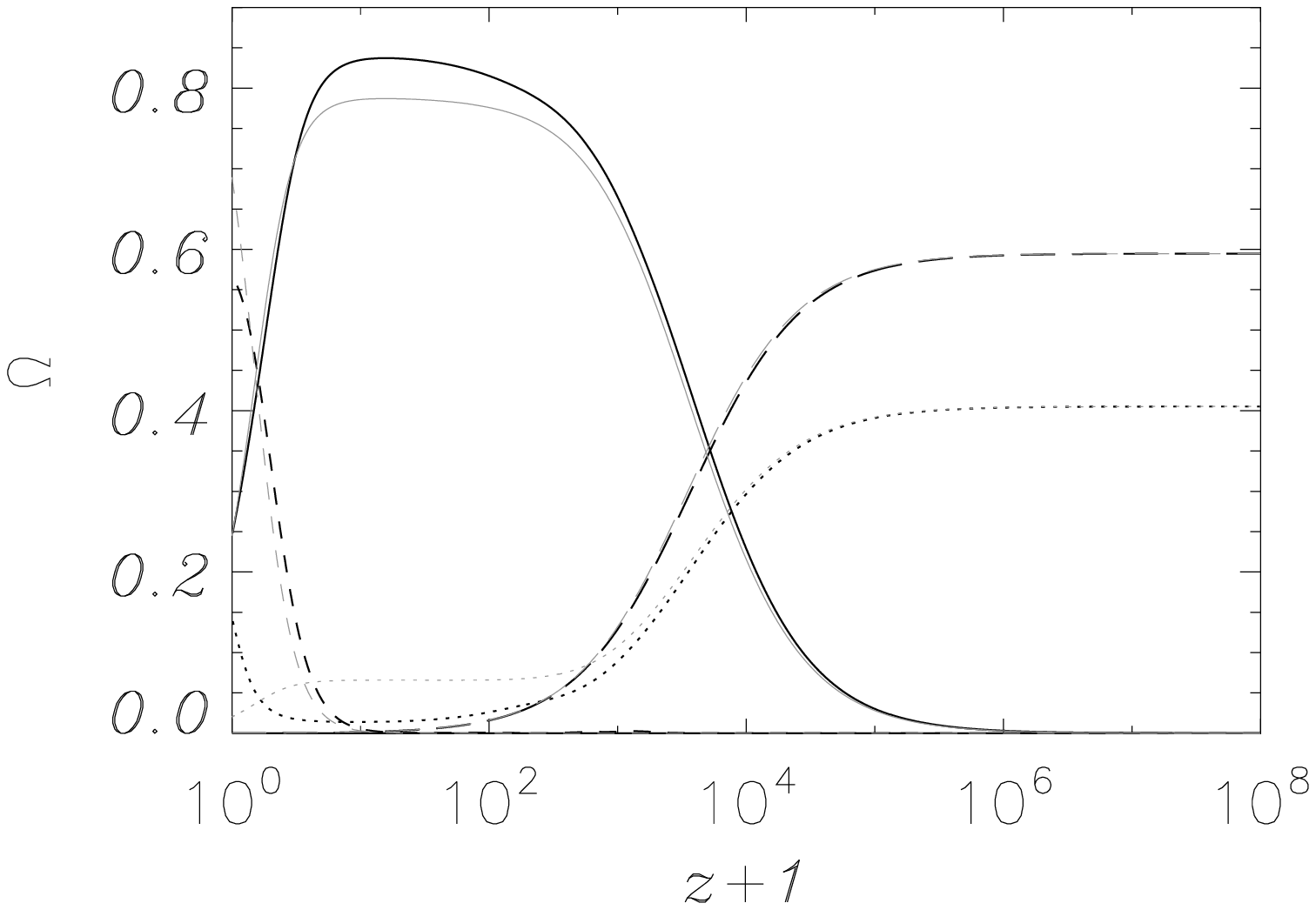}}
\caption{\label{fig:omega_case1}Relative energy densities for Case
  I,II and III.  Solid line: Coupled CDM.  Long-dashed line:
  radiation.  Short-dashed line: dark energy.  Dotted line: neutrinos.
  The faint curves show a typical cosmology with $\beta=0$ for
  comparison.}
\end{center} 
\end{figure} 

\begin{figure*}[t!] 
\begin{center}
\scalebox{0.5}{\includegraphics{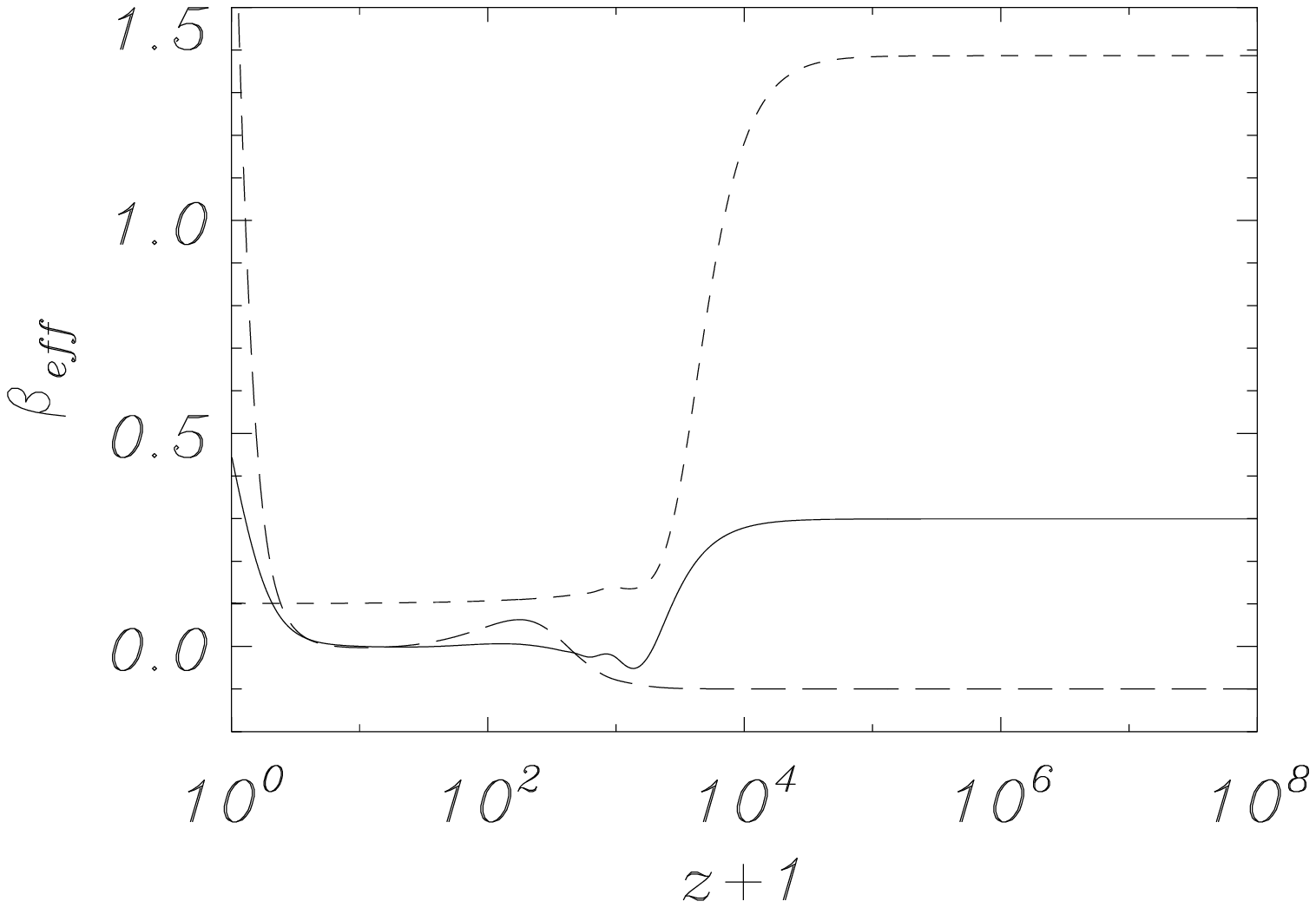}}
\scalebox{0.5}{\includegraphics{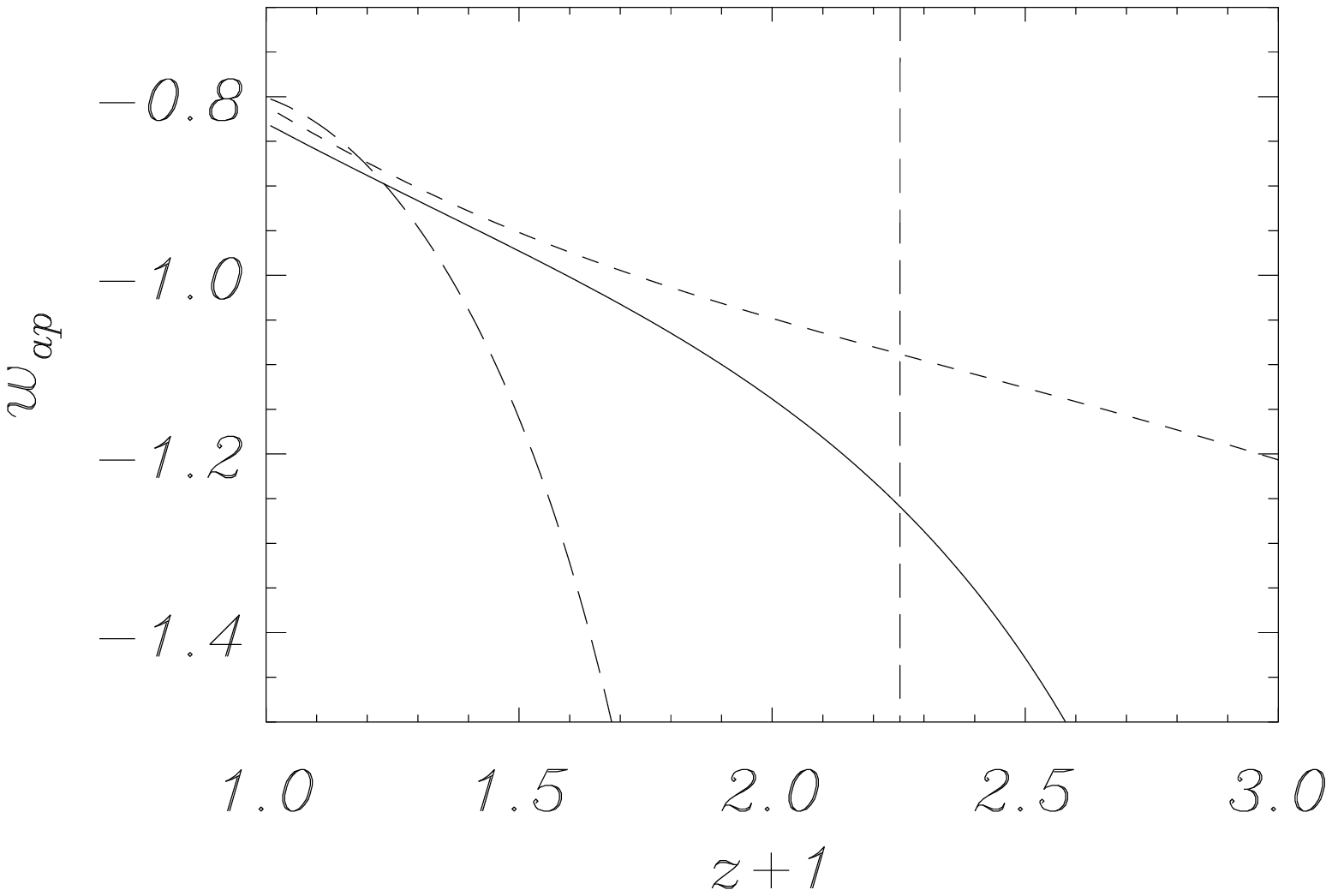}}
\caption{\label{fig:omega_case2}$\beta_{\rm{eff}}$ and $w_{\rm{ap}}$
  for the three cases.  Short-dashed line: Case I.  Solid line: Case
  II.  Long-Dashed line: Case III. Note that for Case III, $x$ (as defined in Eqn.~(\ref{eq:x})) goes through unity when $z\sim 1.2$ and $w_{\rm{ap}}$ becomes undefined.}
\end{center} 
\end{figure*}

The evolution of the density parameters are shown in Fig.~\ref{fig:omega_case1} for the three cases. The effective coupling experienced by the combined fluid, 
as defined in eq. (\ref{eq:betaeff}), is the weighted value of 
the coupling strengths of the two fluids, where the weighting factor is 
given by $(\rho_i - 3p_i)$. Hence, if the fluid with larger coupling 
(e.g. species 2 with $\beta_2$) is dominant at early times, then $\beta_1 \leq \beta_{\rm{eff}} \leq \beta_2$ .  
However, the species with the higher value of coupling also dilutes at a 
greater rate, and so at later times this species becomes sub--dominant, and the 
effective coupling tends towards the lower value for $\beta$. The evolution of $\beta_{\rm eff}$ is 
shown in the left panel of Fig.~\ref{fig:omega_case2}. 

The effect of the couplings on the background evolution is to modify the evolution of the 
energy densities of the coupled fluids. Therefore, even if the density parameters are tuned to the 
standard values today (i.e. $\Omega_\phi\approx 0.7$, $\Omega_{\rm CDM}\approx 0.25$), their values at 
early times are different from what one would expect in the uncoupled case. This will affect, among 
other things, the predictions for the CMB anisotropies. 

As can be seen in Fig.~\ref{fig:omega_case2}, for Case II, the effect
of choosing opposite signs for $\beta$ for the two fluids is to drive
the effective coupling of the combined fluid towards zero during the
matter dominated epoch. Therefore for some period of time the
background cosmology of the system behaves essentially identical to
the usual uncoupled case.  We can
best understand this behavior by considering a transient solution to
Eqns (\ref{eq:Heqn}-\ref{eq:ueqn}) which is valid when the coupled matter species
dominates the energy density.  In this regime we can take $u_i'=0$ and
$x^2,y^2,z^2\ll u_i^2$ and $x',y',z'\sim 0$, which leads to a critical
point given by Eqn (\ref{eq:cp3}), which is equivalent to
$\beta_{\rm{eff}}=0$. This transient regime ends when the energy
density stored in the scalar field becomes non--negligible.

More heuristically, if we instead consider the dynamics of the scalar field, then the
evolution of $\phi$ is driven by the properties of the effective
potential
\begin{equation}
\frac{d V_{\rm{eff}}}{d\phi}\equiv \frac{dV}{d\phi}+\sum_{i=1}^N\beta_i\rho_i,
\end{equation}
which during the matter dominated epoch can be approximated by
\begin{equation}
\frac{d V_{\rm{eff}}}{d\phi} \approx \sum_{i=1}^N\beta_i\rho_i,
\end{equation}
if the potential can be neglected, which is usually the case for quintessence potentials and in models where 
the couplings are of opposite signs. 
Clearly, in this scenario, the effective potential will 
have a minimum located at
\begin{equation}
\phi_{\rm{min}}=\frac{1}{\beta_1-\beta_2}\ln\left(\frac{-\beta_2\rho_{2}^{(*)}}{\beta_1\rho_1^{(*)}}\right).
\end{equation}
We can expect the field to sit in this minimum if $m^2_\phi\equiv \frac{d^2V_{\rm{eff}}}{d\phi^2}>(H/a)^2$, i.e.
\begin{equation}
\left(\frac{H}{a}\right)^2\approx \frac{1}{3}\left(\rho_1+\rho_2 \right)<\beta_1^2\rho_1+\beta_2^2\rho_2,
\end{equation}
which requires
\begin{equation}
\left(\beta_1-\beta_2 \right)\left(\beta_1\beta_2 +1/3\right)>0 \,\,\,\,\, \left(\rm{assuming~~} \beta_2>0\right)
\end{equation}
and hence $\beta_2 > -\frac{1}{3\beta_1}$.
Therefore, provided this constraint is satisfied, and $\beta_1$ and
$\beta_2$ are of opposite sign, we can expect $\beta_{\rm{eff}}$ to be
driven to zero during the matter dominated epoch as the field will be
forced to sit in the minimum of the effective potential.  Note in
particular that this behavior is independent of the {\it form} of $V$,
provided that the scalar field satisfies the conditions discussed above.
When this condition is not satisfied the evolution of the system resembles 
that of Case I.

Case III is unlike the previous two cases in that the values of
$\lambda$ and $\beta$ are chosen such that the system approaches the
accelerated scaling attractor (Attractor B).  This requirement restricts the
choice of density parameter, since $\Omega_i=\Omega_i(\lambda,\beta)$ during the scaling regime (see \cite{Tocchini-Valentini:2001ty}).
As the mass of the growing species is expected to vary by orders of magnitude we choose to
identify the mass-growing species with neutrinos, fully considering the behavior of the neutrinos 
in the relativistic and non--relativistic regimes.  As in Case II, we choose couplings of
opposite sign, such that $\beta_{\rm{eff}}$ is driven towards zero during the matter dominated epoch. 

We compare Case III with the growing-mass model of \cite{Amendola:2007yx}, who consider the dynamics of Attractor B with one coupled species.  
In our scenario, the field sits in the minimum of the effective potential; it becomes heavy due to the presence of multiple contributions from the fluids with couplings of opposite signs.
This is not true in the growing-mass model.  

As a final point we consider the evolution of the apparent equation of
state, as discussed in \cite{Das:2005yj}.  For a system with multiple
species of coupled fluids, the apparent equation of state of the dark
energy sector as measured by an observer assuming an uncoupled system
of fluids is given by
\begin{equation}
w_{\rm ap}= \frac{w_\phi}{1-x},
\end{equation}where $w_\phi\equiv\frac{p_\phi}{\rho_\phi}$ is the
equation of state of the scalar field and
\begin{eqnarray}
x&\equiv& \frac{1}{\rho_\phi}\sum_{i=1}^N\frac{\rho_i^{(0)}}{a^3}\left(1-\frac{\rho_i a^3}{\rho_i^{(0)}}\right) \\
&=& \frac{1}{\rho_\phi}\left(\frac{\rho_c^{(0)}}{a^3}-\rho_c\right). 
\label{eq:x}
\end{eqnarray}
It can be seen that in all of these models $w_{\rm{ap}}$ is rapidly
varying and strongly negative, crossing through $w_{\rm{ap}}=-1$ close to
$z\sim 1$.  
Note that in Case III the apparent equation of state becomes ill-defined, since $x$ goes through unity. This is due to our choice of parameters.

\section{Perturbations}
\label{sec:perturbations}
In the previous section we demonstrated that the
background evolution of a cosmology with two fluids coupled to a
single scalar field could be mimicked by a system with a single fluid
coupled to a scalar field with an effective coupling given by
eqn. (\ref{eq:betaeff}). It is therefore impossible to distinguish
between the two systems using probes of the cosmological background,
such as measuring the expansion history. We have also seen that a
system which comprises of multiple fluids with constant coupling
strengths $\beta_i$ may appear to be a single fluid with a time
varying $\beta$, or indeed a system for which $\beta=0$ for at least
some part during the matter dominated era.  We now investigate the
effect that these couplings have on the evolution of cosmological
perturbations, and determine whether the degeneracy between the single
fluid and the multi--fluid approach can be broken.

For each individual fluid, the equations governing the evolution of
the density perturbations can be obtained from the perturbation of
eqn.~(\ref{eq:tmumu}) and, at first order, are given in the
synchronous gauge by (see e.g. \cite{Rhodes:2003ev,Amendola:2003wa})
\begin{eqnarray}
\label{eq:perteqns}
\dot{\delta}_i &=& 3\left(H+\beta_i
\dot{\phi}\right)\left(w_i-\frac{\delta p_i}{\delta
  \rho_i}\right)\delta_i - \left(1+w_i\right)\left(\theta_i +
\frac{\dot{h}}{2}\right)\nonumber \\ &+&
\beta_i\left(1-3w_i\right)\dot{\delta
  \phi}+\frac{d\beta_i}{d\phi}\dot{\phi}\delta\phi\left(1-3w_i\right)\\
\dot{\theta_i} &=& -H(1-3w_i)\theta_i -
\frac{\dot{w_i}}{1+w_i}\theta_i+\frac{\delta p_i / \delta
\rho_i}{1+w_i}k^{2}\delta_i \nonumber \\ &+&
\beta_i\frac{1-3w_i}{1+w_i} k^{2} \delta
\phi-\beta_i(1-3w_i)\dot{\phi}\theta_i - k^{2} \sigma_i, 
\end{eqnarray}
where for each species $i$ we have defined: $\delta_i = \delta
\rho_i/\rho_i$ is the density contrast, $\delta p_i$ is the pressure
perturbation, $h$ is a the metric perturbation, $\theta_i$ is the
gradient of the velocity field, $\sigma$ is the shear stress and $\delta \phi$ is the perturbation
in the scalar field.
For the HDM case, we use the full Boltzmann hierarchy to evaluate the density perturbations~\cite{Brookfield:2005bz,Ichiki:2007ng}.

The perturbed Klein-Gordon equation, governing the evolution of the
scalar field perturbation $\delta \phi$ is given by~\cite{Rhodes:2003ev,Amendola:2003wa}
\begin{eqnarray}\label{pertkg}
\lefteqn{\ddot{\delta \phi}+ 2H \dot{\delta \phi}+\left(k^{2} +
a^{2}\frac{d^{2}V}{d\phi^{2}}\right)\delta \phi+\frac{1}{2}\dot{h}
\dot{\phi}=} \\ \nonumber & &-a^2 \sum_{i=1}^N \left[\beta_i(\delta\rho_i-3\delta p_i)+\frac{d\beta_i}{d\phi}\delta\phi(\rho_i-3 p_i)\right].
\end{eqnarray} 
The evolution of the density contrasts in the different cases are
shown in Fig.~\ref{fig:densitycontrast}, where we consider the
limits of large and small scales. We now discuss both regimes separately. 

\subsection{Small Scales}
\begin{figure*}[t]
\begin{center}
\subfigure[Small Scales]{\label{fig:densitycontrast_b}
\begin{minipage}[h]{.45\textwidth}
\scalebox{0.5}{\includegraphics{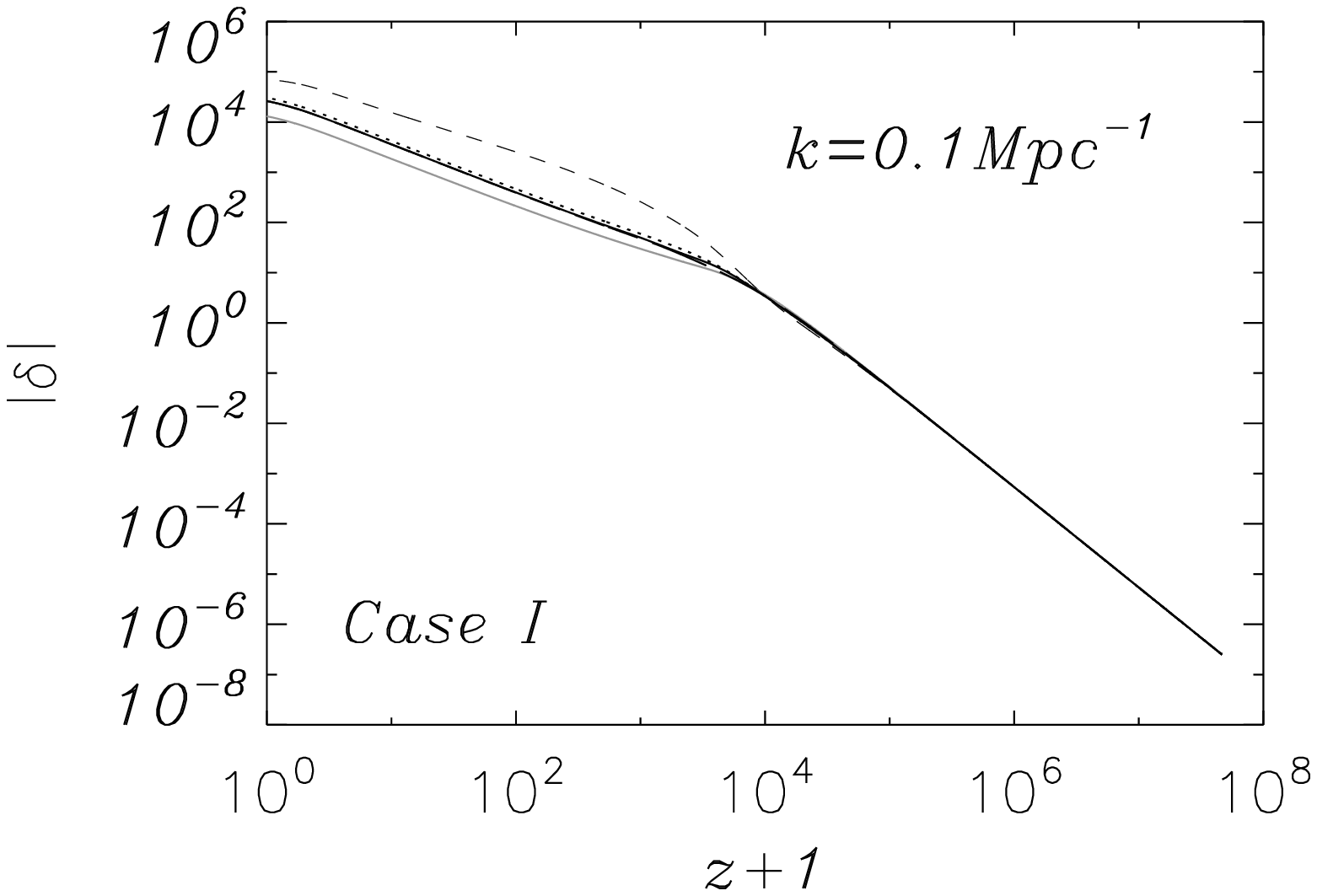}}
\scalebox{0.5}{\includegraphics{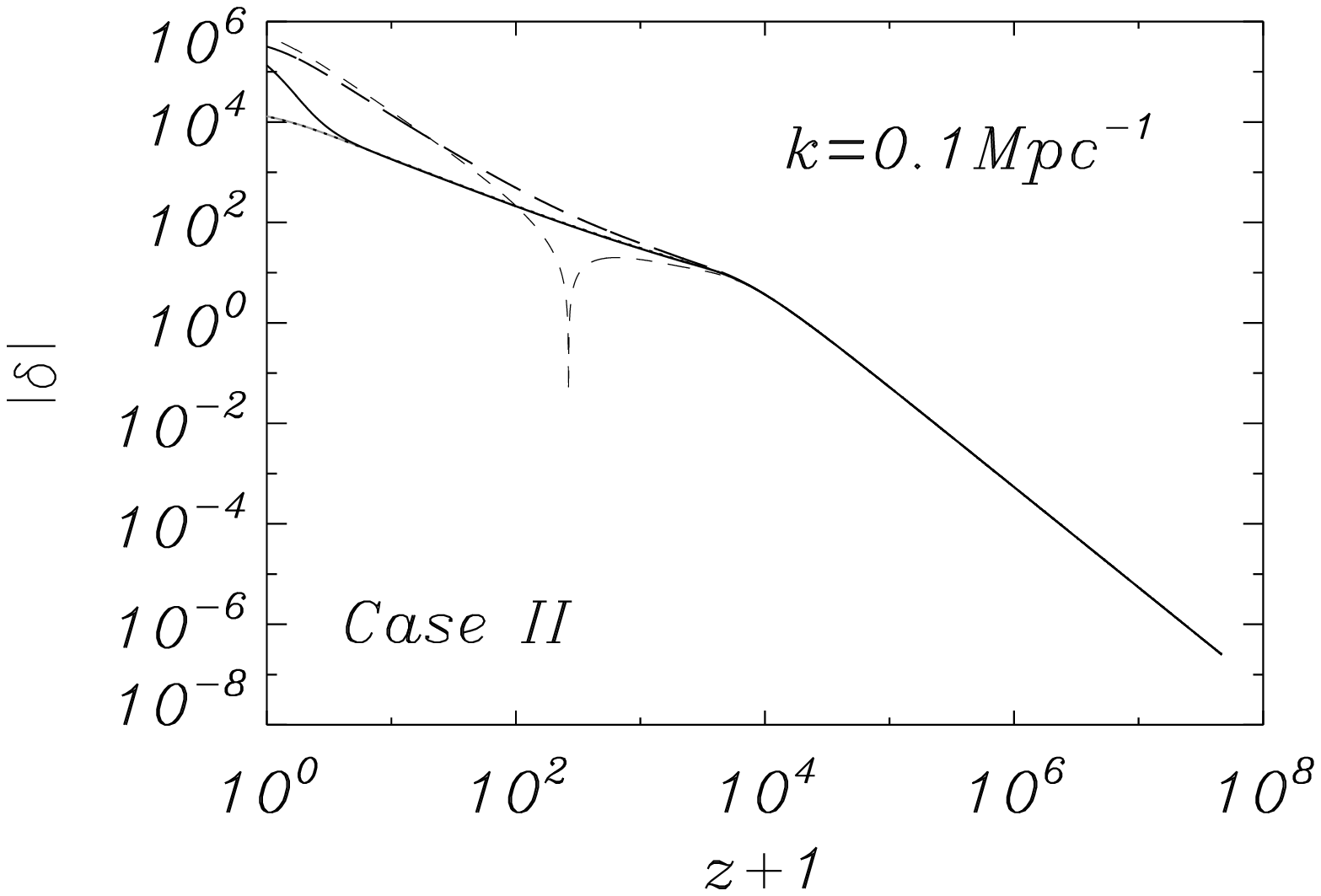}}
\scalebox{0.5}{\includegraphics{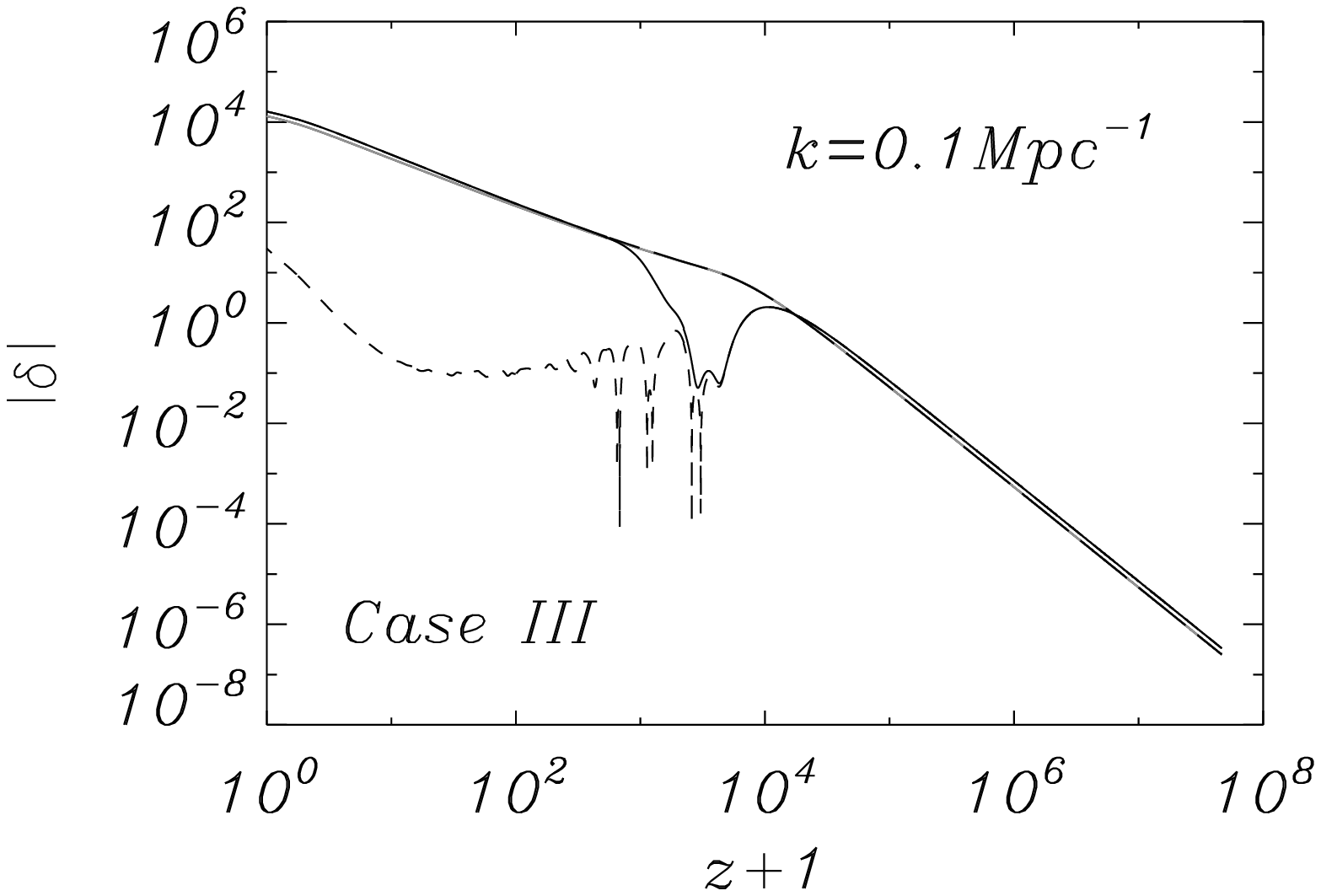}}
\end{minipage}
}
\subfigure[Large Scales]{\label{fig:densitycontrast_a}
\begin{minipage}[h]{.45\textwidth}
\scalebox{0.5}{\includegraphics{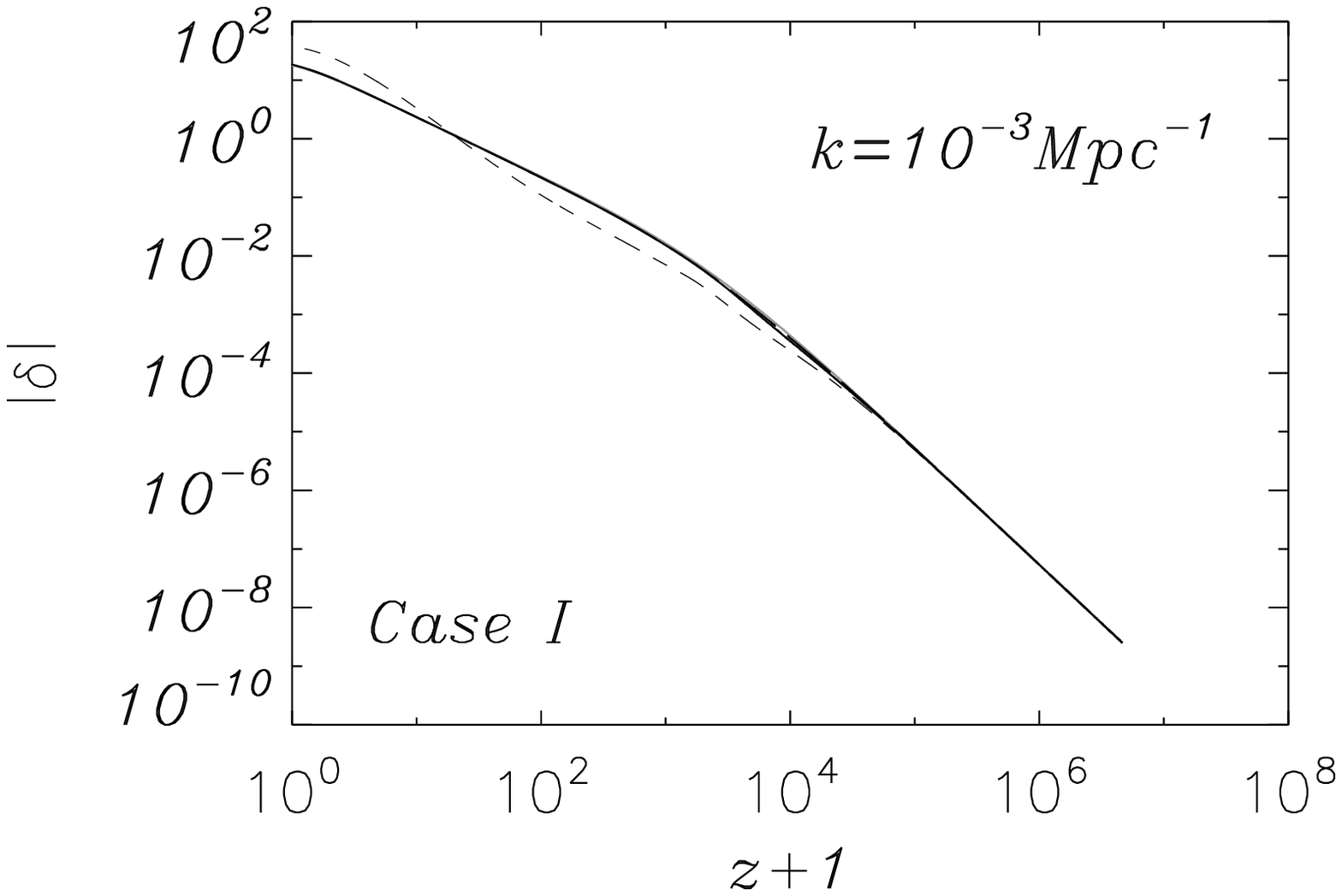}}
\scalebox{0.5}{\includegraphics{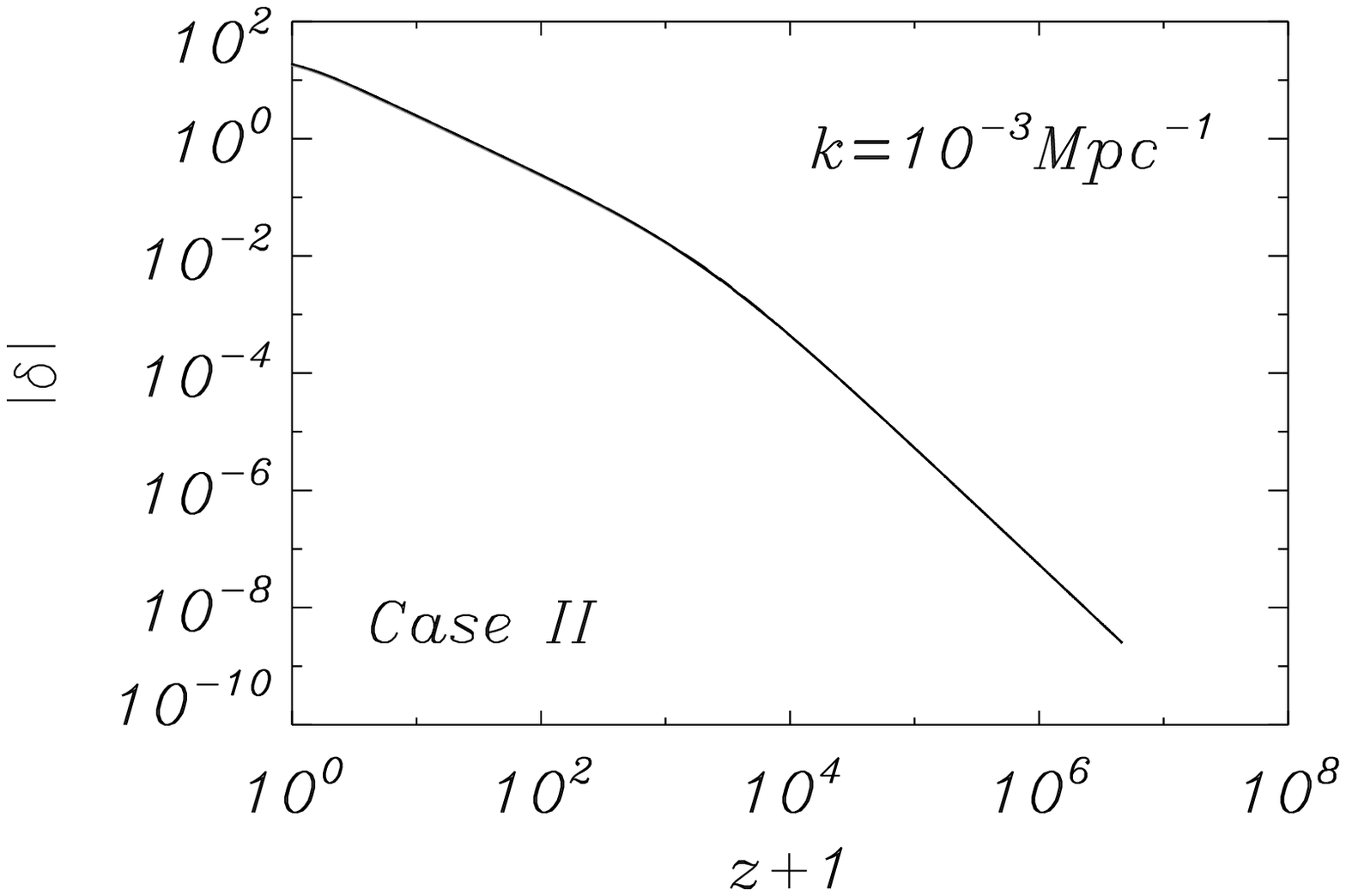}}
\scalebox{0.5}{\includegraphics{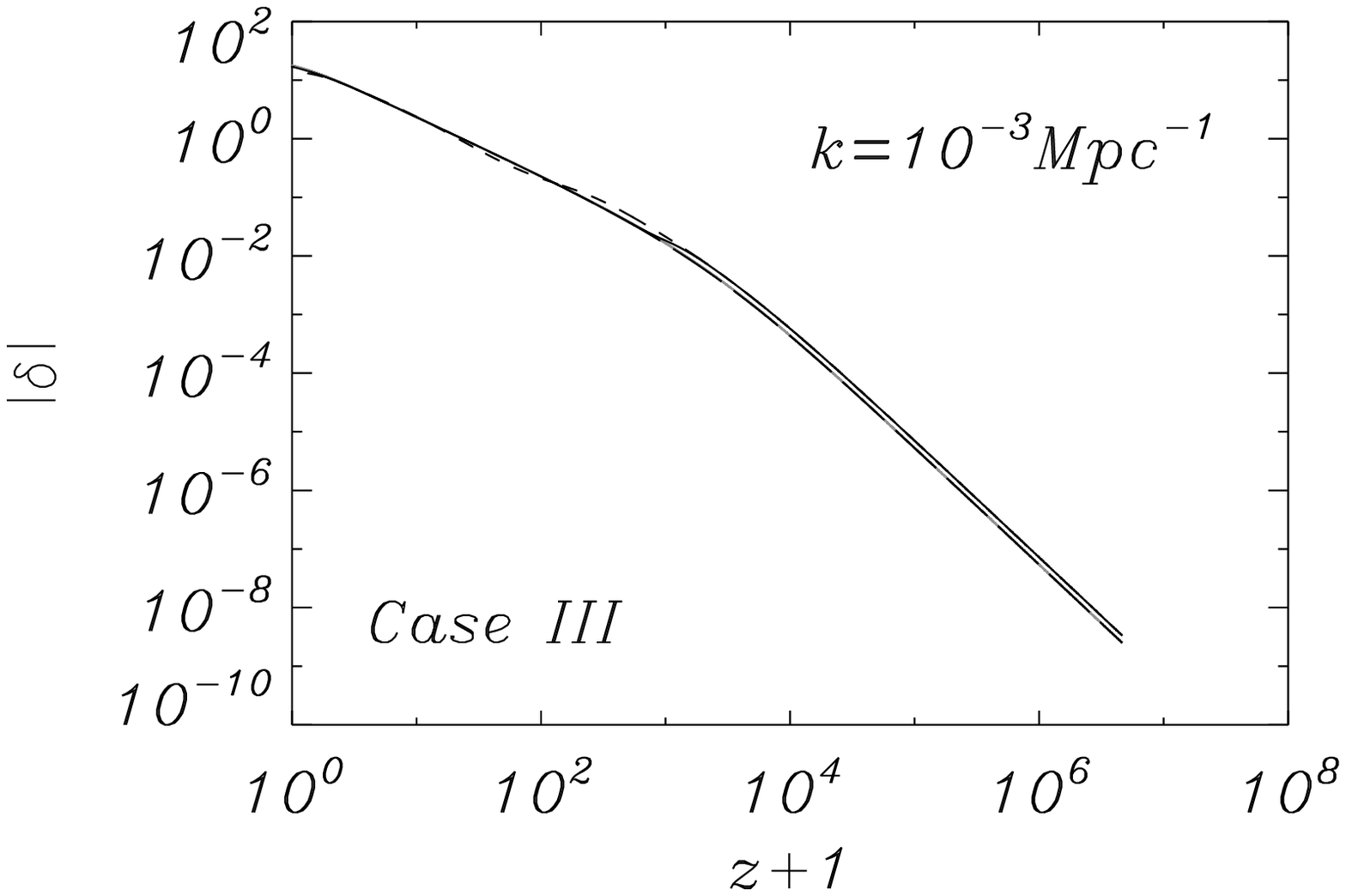}}
\end{minipage}
}
\caption{\label{fig:densitycontrast} Evolution of the density
  contrasts for the three cases on small and large scales. Thin 
  solid line: $\beta=0$.
  Short dashed line: $\delta_1$.  Long dashed line: $\delta_2$.  Thick
  solid line: $\delta_c$.  Dotted line (Cases I and II only): equivalent single fluid with
  $\beta=\beta_{\rm{eff}}$.}
\end{center}
\end{figure*}
\begin{figure*}[t] 
\begin{center}
\subfigure[Matter Power Spectrum]{\label{fig:powerspectra_a}
\begin{minipage}[h]{.45\textwidth}
\scalebox{0.5}{\includegraphics{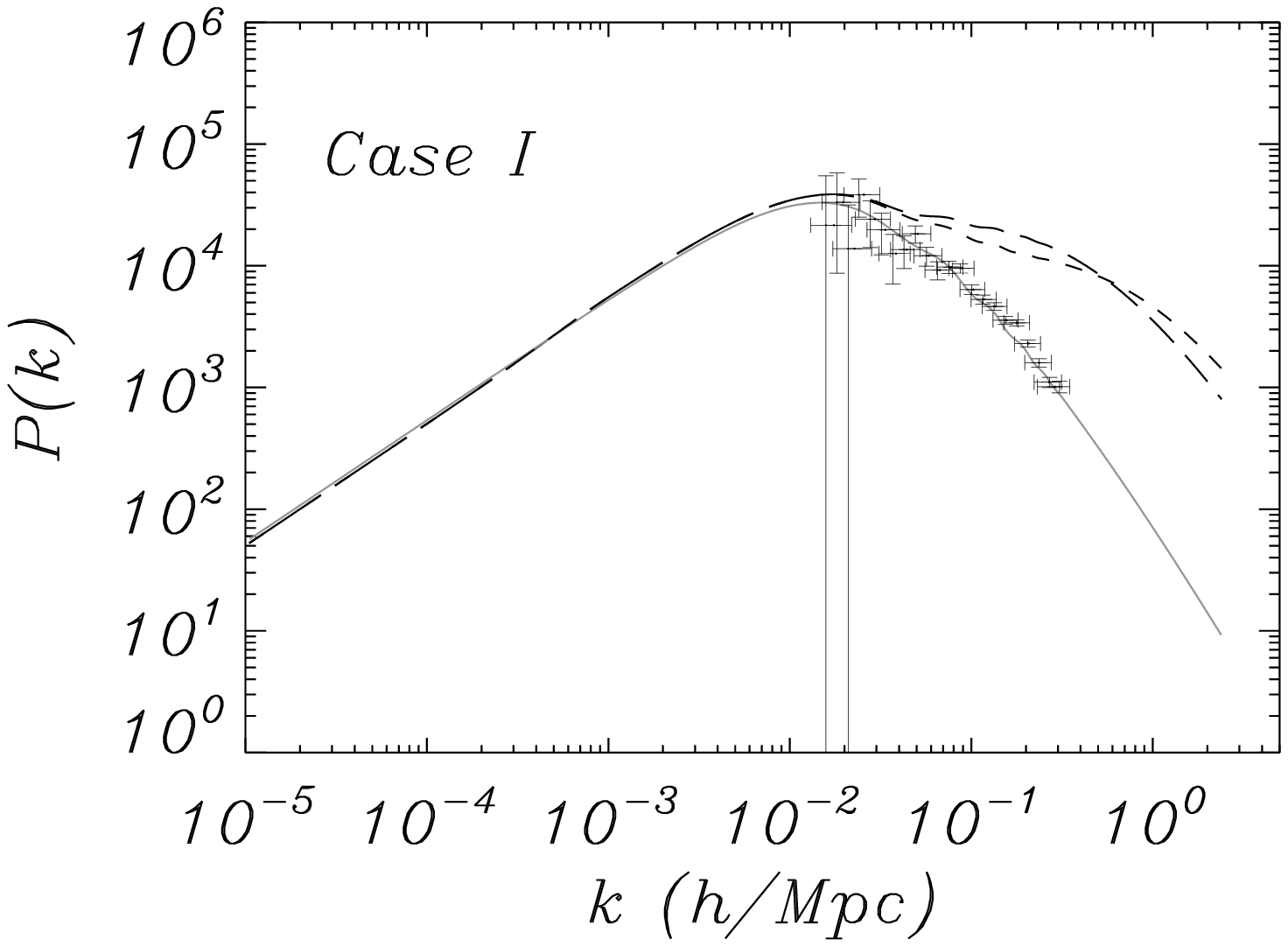}}
\scalebox{0.5}{\includegraphics{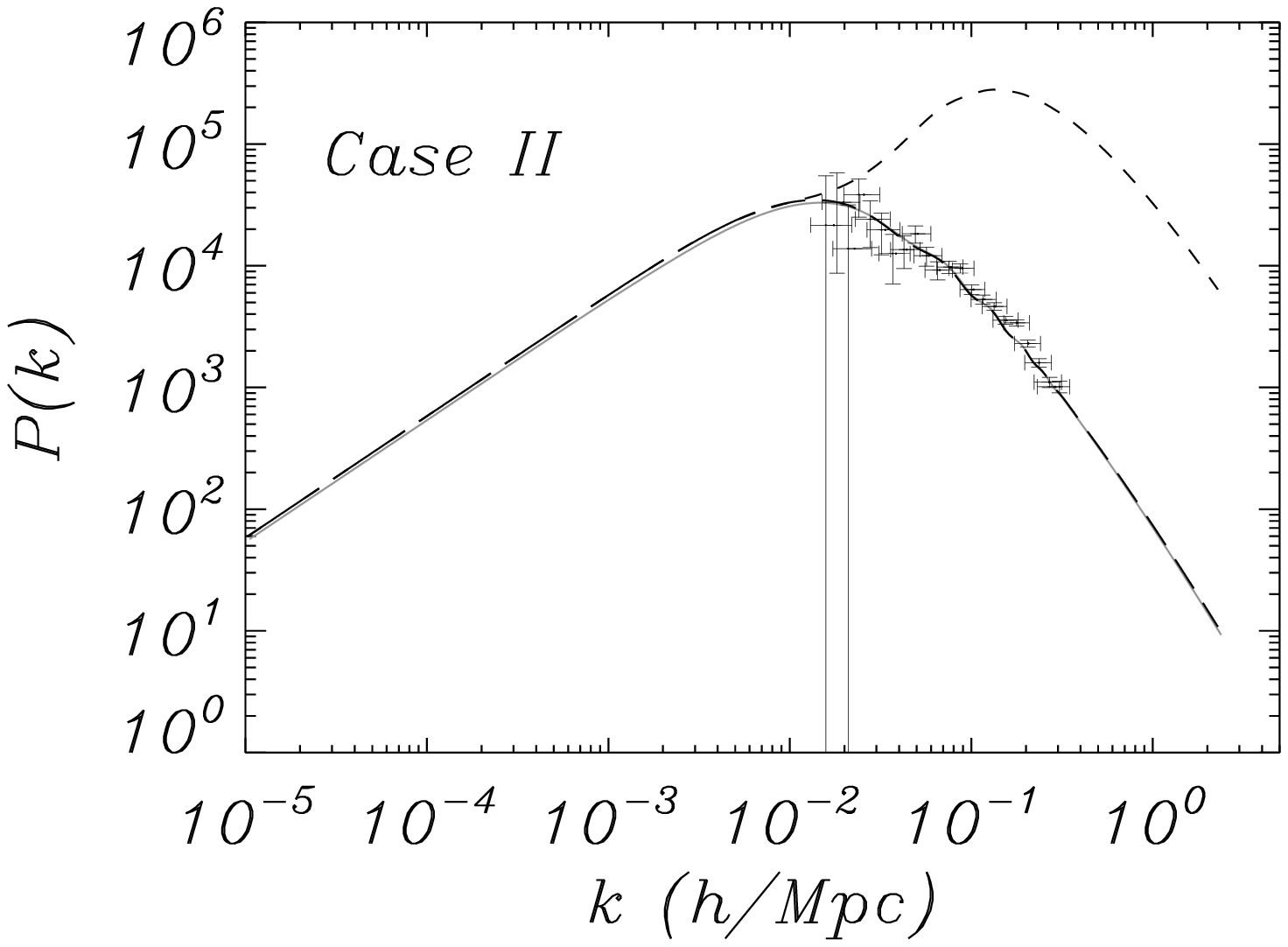}}
\scalebox{0.5}{\includegraphics{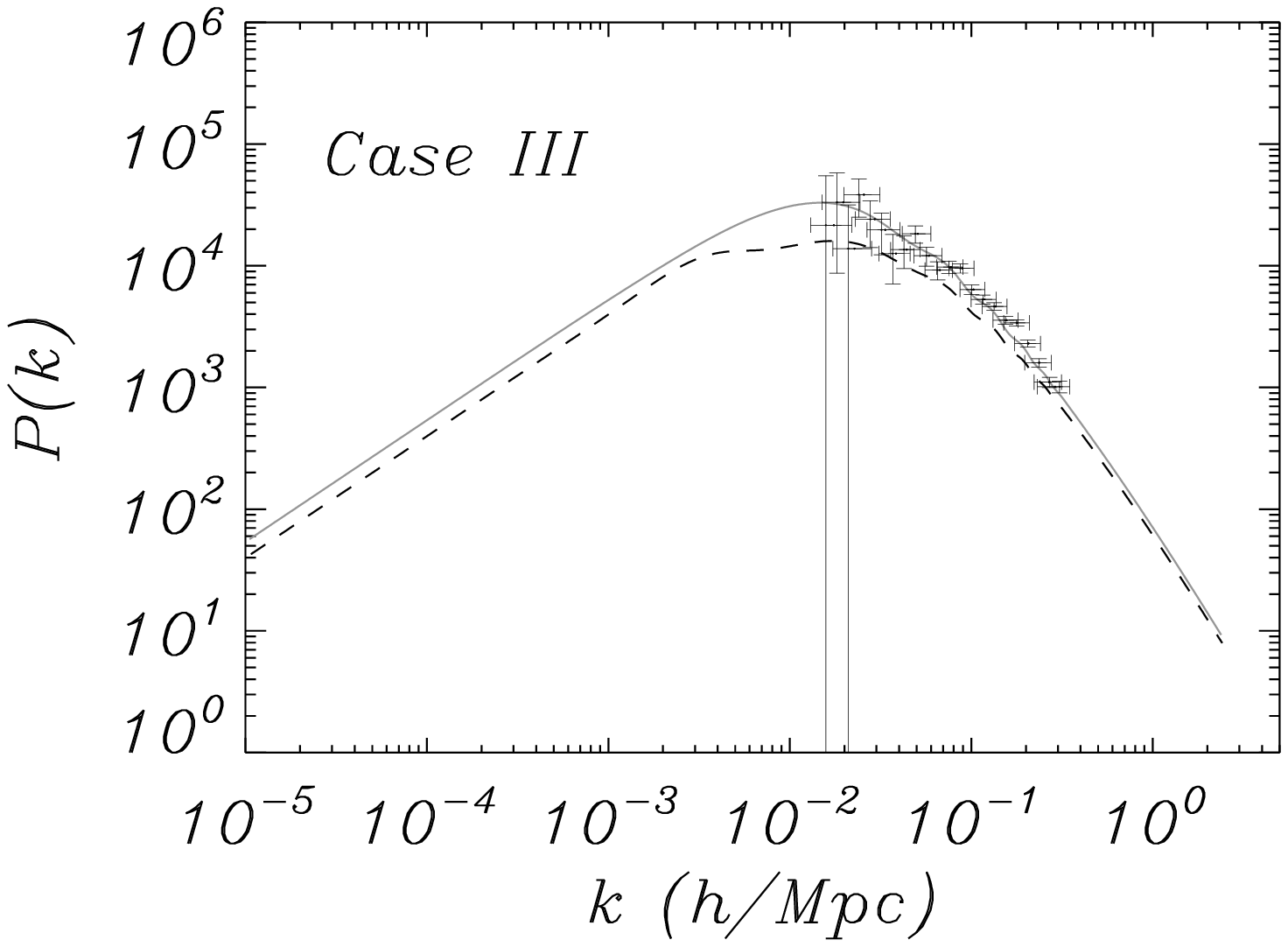}}
\end{minipage}
}
\subfigure[Temperature Anisotropy Spectrum]{\label{fig:powerspectra_b}
\begin{minipage}[h]{.45\textwidth}
\scalebox{0.5}{\includegraphics{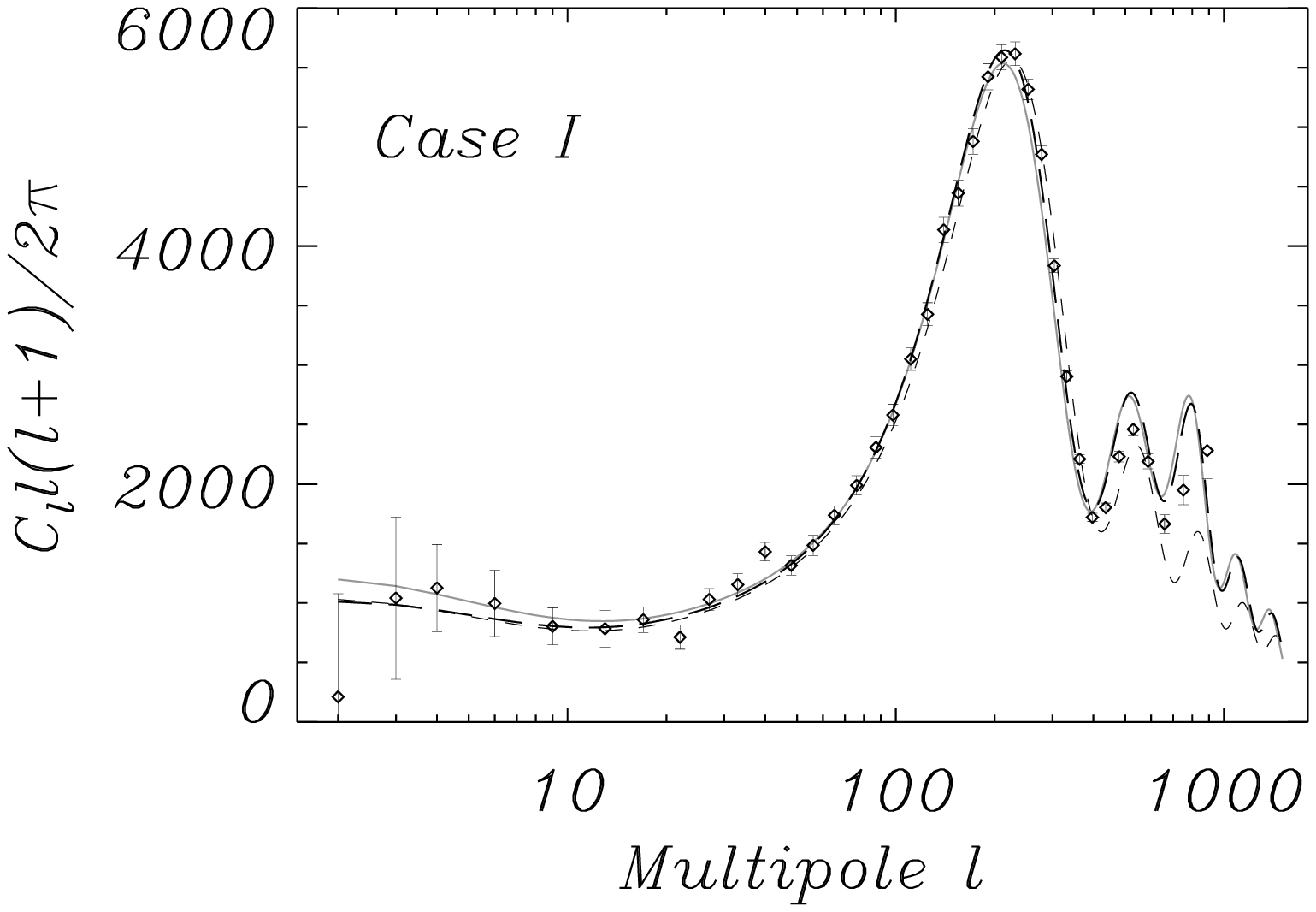}}
\scalebox{0.5}{\includegraphics{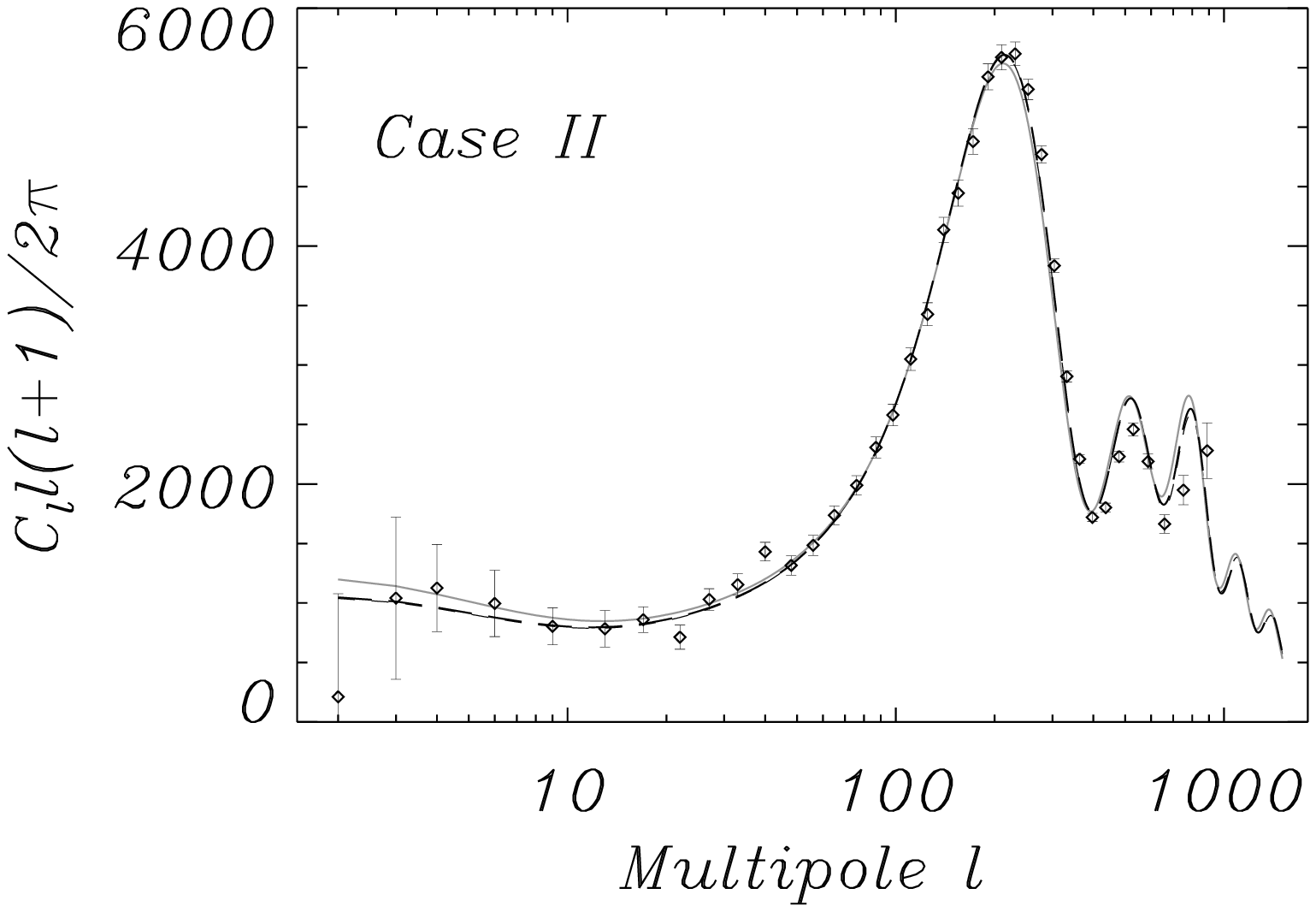}}
\scalebox{0.5}{\includegraphics{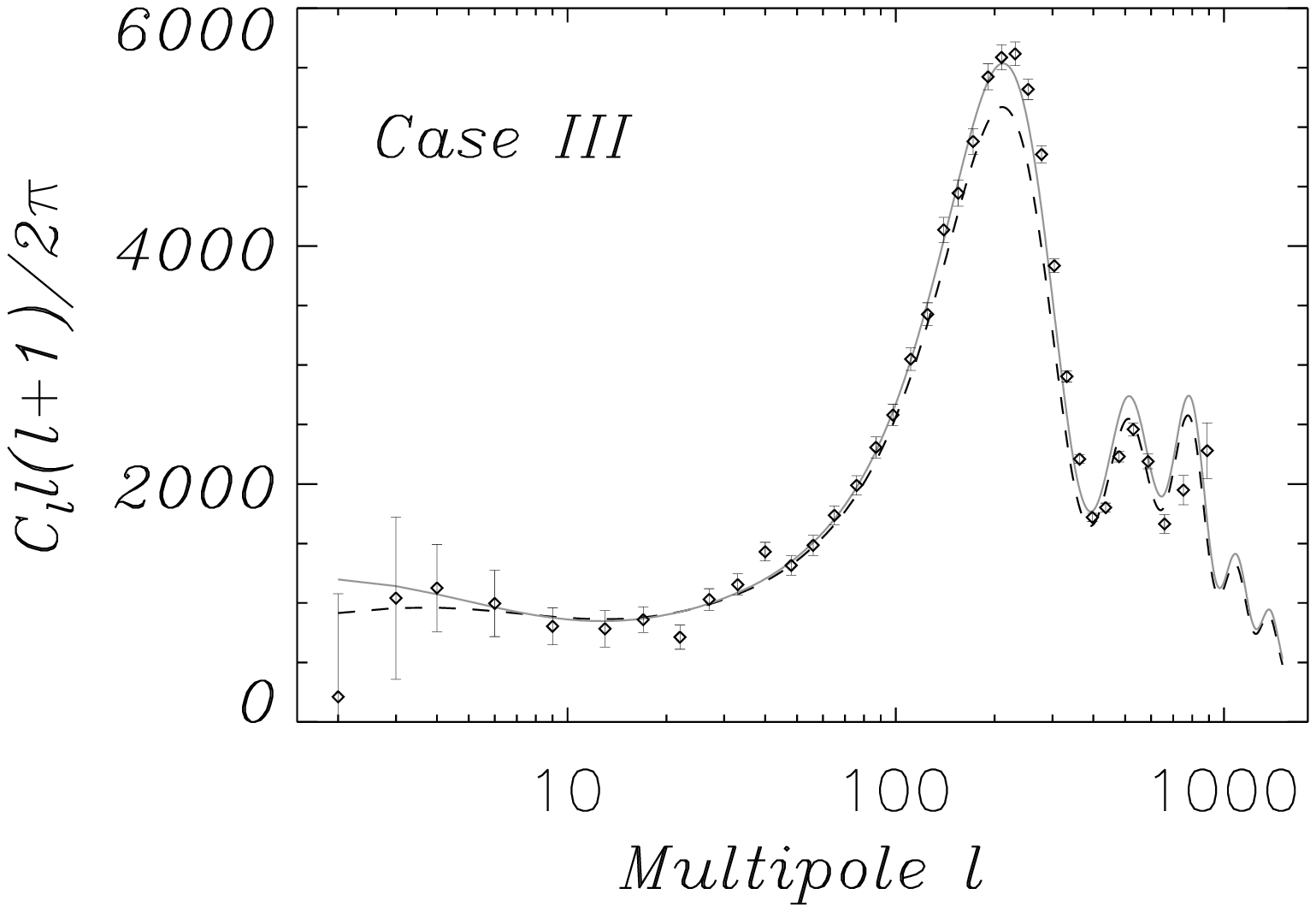}}
\end{minipage}
}
\caption{\label{fig:powerspectra} CMB anisotropy spectra and matter
  power spectra for the three cases. Solid line: $\beta=0$. Short dashed line: 2 coupled fluids.  Long dashed line (Cases I and II only): single fluid with $\beta=\beta_{\rm{eff}}$.}
\end{center} 
\end{figure*} 
On small scales, the perturbed Klein Gordon equation can be
approximated by
\begin{equation}
k^2 \delta\phi +\frac{1}{2}\dot{h}\dot{\phi} = -a^2
\sum_{i=1}^N\beta_i \delta\rho_i,
\end{equation}
where we have assumed that $\beta_i$ is constant and $w_i\sim 0$. Using one 
of the perturbed Einstein equations \cite{Ma:1995ey}
\begin{equation}
\ddot{h}+H\dot{h}=-a^2\sum_j \delta \rho_i
\end{equation}
we can show that
\begin{equation}
\ddot{\delta_i}=-\left(H+\beta_i \dot{\phi} \right)\dot{\delta_i}+ \frac{3}{2}H^2
\sum_{j=1}^N \Omega_j\delta_j \frac{G_{ij}^{\rm{eff}}}{G_N}
\end{equation}
where
\begin{equation}
G_{ij}^{\rm{eff}}=G_{\rm{N}}\left(1+2\beta_i\beta_j\right).
\end{equation}
This equation has also been obtained in \cite{Amendola:2003wa}.

In Case I, where $\beta_1\beta_2>0$, the two species experience an
additional attractive force between both similar and dissimilar particles,
increasing the growth of the density perturbations.  However in Case
II, $\beta_1$ and $\beta_2$ have opposite signs, so whilst like particles feel an additional attractive force, two dissimilar particles
experience a repulsive force between each other.
This effect can be observed in the evolution of the density contrast
in Fig.~\ref{fig:densitycontrast_b}; $\delta_i$ is enhanced in Case I whilst in Case II $\delta_2$  is enhanced but $\delta_1$ is supressed (note that $\delta_1$ becomes negative).  For Case II, after horizon crossing, the dominant
fluid (species 2) clusters at an enhanced rate due to the additional
contribution to $G^{\rm{eff}}$. Species 1 also feels an additional
attractive force towards itself, but also experiences a repulsive
force from species 2. As species 2 dominates the energy density, the
bulk of the ordinary matter falls into the gravitational potential
wells that it generates, whilst species 1 is repulsed from these
overdensities.

However, note that during the period when the field is in it's
minimum, the behavior of the combined fluid is identical to that of a
single uncoupled species. It is only at late times, when the field
evolves away from it's minimum due to the onset of dark energy
domination, that the extra coupling force is felt, and the
perturbations in the total fluid grow at an enhanced rate.  

Case III is similar to Case II, except that the HDM undergoes freestreaming until it becomes non-relativistic, which suppresses the growth of the density contrast ($\delta_1$).

\subsection{Large Scales}
On superhorizon scales the evolution of $\delta_i$ can be approximated by
\begin{equation}
\dot{\delta_i}=-\frac{\dot{h}}{2}+\beta_i\dot{\delta\phi}
\label{delta_super}
\end{equation}
for $w_i=0$, which reduces to the usual expression for small values of coupling.
The suppression of growth (prior to horizon crossing, $z+1\geq 10^2$), seen in Fig.~\ref{fig:densitycontrast_a} for
the highly coupled species, arises due to this additional coupling 
term, as the growth of $\delta_i$ is modified by the term $\beta_i\delta\phi$.  In our models, $\delta\phi$ is negative and we observe a suppression of $\delta$. 
However, once the mode enters the horizon, Eqn.~(\ref{delta_super}) is no longer valid, the additional force between the particles 
mediated by dark energy can be felt and the growth of $\delta_i$ is enhanced.

We point out that for all of the models considered in this
paper, the range of the additional scalar force
\begin{equation}
\lambda \sim \frac{1}{m_\phi}, {\rm where}\,\,\, m^2_\phi \equiv \frac{d^2V}{d\phi^2}
\end{equation}
is much greater than the horizon size and we do not therefore observe any 
effects due to the finite range of the force.  

\subsection{Comparison to Equivalent Single Fluid}
As for the background, we can consider the two species of matter to be a single fluid with
\begin{align}
\delta\rho_c=\sum_{i=1}^N\delta\rho_i, \qquad \delta_c=\frac{\delta\rho_c}{\rho_c}, \\
\left(\rho_c+p_c \right)\theta_c=\sum_{i=1}^N\left(\rho_i + p_i\right)\theta_i.
\end{align}
As explained in Section \ref{sec:background}, an observer studying the
background evolution, who assumes only a single fluid dark matter
sector, will infer that the coupling evolves in time, $\beta_{\rm
  eff}=\beta_{\rm eff}(\phi)$.  With two dark matter species, an
observer might detect a coupling of the form given in
Eqn.~(\ref{eq:singlebetaeff}).

However, it is apparent that we cannot re-write Eqns
(\ref{eq:perteqns}) and (\ref{pertkg}) into an equivalent single fluid
form, due to the $\frac{d\beta}{d\phi}$ term.  Whilst for the single
fluids $\beta$ is purely a function of $\phi$, in the combined fluid
$\beta$ becomes a function of $\phi$ and $\rho_i$.  The term
$\delta\beta$ receives contributions from the variation of $\phi$ and
the variation of $\delta\rho_i$, which itself is a function of $\phi$
and $h$.  Hence, using $\beta_{\rm eff}(\phi)$ (inferred from
background observations) to calculate the perturbation evolution would
lead to incorrect predictions for $\delta$.  For example, assuming a
single fluid with an effective coupling leads to a different evolution
of perturbations than if the observer had correctly identified two
species.

In order to highlight this observational difference (allowing us to
break the degeneracy), we calculate the evolution of perturbations for
an equivalent single fluid model with field-varying coupling
$\beta_{\rm eff}(\phi)$. This is plotted in
Fig.~\ref{fig:densitycontrast} for a direct comparison for Cases I and
II.  (This effect is most apparent when we consider the matter power
spectrum in the next section.) We also plot the case $\beta=0$ for
comparison.

In Case II, when $\beta_{\rm eff}=0$, the density contrast of the
combined fluid, $\delta_c$, mimics a single uncoupled fluid, which of
course coincides with the equivalent single fluid (with $\beta_{\rm
  eff}=0$) during this time.

For Case III, however, we do not compare the two fluid case with an
equivalent single fluid.  This is because $\beta_{\rm{eff}}$ is a
function of both the pressure and energy density of the coupled
species, one of which is highly relativistic at early times.  Although
$\beta_{\rm{eff}}$ could be reconstructed as $\beta_{\rm eff}(\phi)$,
it would not take the simple analytic form given in
Eqn.~(\ref{eq:singlebetaeff}).

\subsection{Observational Signatures}
The differing behaviour of the perturbations for the multiple
fluid and the single effective fluid leaves a detectable imprint on
both the anisotropy spectrum and the matter power spectrum, as shown in 
Fig.~\ref{fig:powerspectra}. 
Once again, we compare the results for the multiple coupled fluids
with the equivalent single coupled fluid with $\beta=\beta_{\rm{eff}}(\phi)$.
We remind the reader that the parameters are chosen such that the new effects of the interactions are emphasized.

For both Cases I and II, the matter power spectra are relatively
unchanged on large scales, as the growth of density perturbations is
weakly affected by the coupling (see
Fig.~\ref{fig:densitycontrast_b}).  On small scales, however, the
enhanced growth of $\delta_c$ is seen explicitly in the matter power
spectra.  The difference between multiple and single field fluids is
apparent and breaks the degeneracy seen in the background solution.

In Case III, we see a suppression of power in the matter spectrum
compared with the standard case.  This is a consequence of the mass
variation of the neutrinos, which are lighter in the past than in the
standard case.  They therefore undergo freestreaming for longer, which
suppress the growth of density perturbations.  This effect is partly
compensated by the increased effective Newton's constant, $G_{ij}^{\rm
  eff}$, experienced by the CDM particles, which would otherwise clump
more than in the standard case.

Fig.~\ref{fig:powerspectra_b} shows the effect of the coupled fluids
on the temperature anisotropy spectrum.  It has been shown previously
that for a single coupled species of CDM, the coupling is constrained
to $\vert\beta\vert\lesssim 0.1$~\cite{Amendola:2003eq}, due to the
large contribution from the coupling to the late--time Integrated
Sachs--Wolfe (ISW) effect.  For Cases I and II, however, we observe a
reduction in the ISW effect, even though we have coupling strengths
$O(1)$.  This is because the effective coupling strength of the
combined fluid in both Case I and Case II has been chosen to be
$O(0.1)$ at late times, to ensure that the ISW effect does not
contribute too greatly on large scales.  It should also be noted that,
in a system with multiple coupled fluids, the fluid with the smallest
$\beta$ dominates the energy density of the coupled species at late
times, as this species dilutes slower than species with larger
couplings.  In this scenario, therefore, we are not resticted by
previous constraints on $\beta$.

For Case III, we can see that the contribution to
the ISW from the coupling is enhanced relative to the height of the
first acoustic peak.  

An additional signature can be found in comparing the relative heights
of the second and third peaks, which are significantly altered in Case
I.  This arises partly due to the change in relative energy densities
at the time of last scattering as seen in Fig.~\ref{fig:omega_case1}.
For Cases II and III, the relative energy densities at this time
remain unchanged as $\beta_{\rm{eff}}\rightarrow0$ and the fluids
mimic standard cosmology.

We can also see that there is a significant variation in the
anisotropy spectrum and the matter power spectrum for the two coupled
species of matter in comparison with the equivalent single species of
matter.  This difference therefore leaves a detectable signature,
which allows differentiation between coupled fluids of single or
multiple species.

\section{Conclusions}
\label{sec:conclusions}
In this paper, we have considered the effect of new forces mediated by
dark energy, considering a multi-component dark matter sector. 
From our study of the background, we have shown
that multiple coupled fluids can mimic a single coupled species.  In
particular, multiple fluids with constant couplings (well motivated in
particle physics) can behave like a single fluid with a time varying
coupling.  Additionally, the cosmological background can behave like a
$\Lambda$CDM universe during the matter dominated epoch, for models in
which the effective coupling vanishes.

We established that perturbations allow us to break this degeneracy
between a single effective dark matter species and multiple fluids.
Specifically, observers could distinguish between a single fluid with
a time varying coupling $\beta(\phi)$ and an effective fluid with an
apparent time varying coupling.

Perturbation growth is affected on all scales by the couplings.  On
scales smaller than the horizon, particles feel a new force, mediated
by the scalar field. Depending on the relative sign of the coupling,
this force can be either attractive or repulsive.  Depending on the
coupling strength and relative densities, the growth of perturbations
for a given species can be either enhanced or suppressed. We remark that 
the instabilities observed in \cite{Bean:2007nx,Bean:2007ny} are not present in the models 
considered here. 

Compared to previous studies with a single fluid, CMB constraints on
the strength of multiple couplings can be relaxed.  We find that the
system is not constrained to have $\beta_i\lesssim 0.1$.  We expect
stronger constraints to come from large scale structure surveys
probing the matter power spectrum.  Additionally, supernovae
observations will constrain the evolution of the apparent equation of
state, which can significantly deviate from $w=-1$ and can even be
$w\leq -1$. Observational constraints on these models should be
considered in future work. Additionally, a study of non-linear perturbations 
on small scales should be undertaken, to identify further signatures of 
these interactions (such as in \cite{Nusser:2004qu} and \cite{Maccio:2003yk}).
For example, the new interactions could modify the density profiles of dark haloes 
or induce a bias between baryons and dark matter.  For a single coupled species, this was studied in \cite{Maccio:2003yk}, who conclude that cosmologies with dark-dark couplings are viable from an observational standpoint.
It would be interesting to generalise these findings to our models, but this is outside the scope of this paper. 

\section*{Acknowledgments}
We are grateful to David Mota, Domenico Tocchini--Valentini and
Christof Wetterich for useful discussions.  The authors acknowledge
support from STFC.

\bibliographystyle{h-physrev3}
\bibliography{awb}

\end{document}